\documentclass[aps,prd,groupedaddress,nofootinbib,twocolumn]{revtex4-2}
\usepackage{graphicx,color,amsmath,mathtools,amssymb}
\usepackage{tcolorbox}
\usepackage{amsthm}
\usepackage{graphicx}
\usepackage{color,soul}
\usepackage{soul}
\usepackage{bm}
\usepackage{subfigure}
\usepackage{multirow}
\usepackage{hyperref} 
\usepackage{slashed}
\hypersetup{colorlinks=true,citecolor=blue,linkcolor=magenta,filecolor=magenta,urlcolor=blue,}

\begin{document}
	\title{ $Z_{cs}$, $Z_c$ and $Z_b$ states under the complex scaling method}
	
	\author{Jian-Bo Cheng$^{1}$}%
	\email{jbcheng@pku.edu.cn}
	\author{Bo-Lin Huang$^{1}$}%
	\email{blhuang@pku.edu.cn}
	\author{Zi-Yang Lin$^{1}$}
	\email{lzy\_15@pku.edu.cn}
	\author{Shi-Lin Zhu$^{1}$}
	\email{zhusl@pku.edu.cn}
	\affiliation{
		$^1$School of Physics and Center of High Energy Physics, Peking University, Beijing 100871, China\\
	}%
	
	\date{\today}
	
	\begin{abstract}
		
		We investigate the $Z_b$, $Z_c$ and $Z_{cs}$ states within the
		chiral effective field theory framework and the $S$-wave single
		channel molecule picture. With the complex scaling method, we
		accurately solve the Schr\"odinger equation in momentum space. Our
		analysis reveals that the $Z_b(10610)$, $Z_b(10650)$, $Z_c(3900)$
		and $Z_c(4020)$ states are the resonances composed of the $S-$wave
		$(B\bar{B}^{*}+B^{*}\bar{B})/\sqrt{2}$, $B^{*}\bar{B}^*$,
		$(D\bar{D}^{*}+D^{*}\bar{D})/\sqrt{2}$ and $D^{*}\bar{D}^*$,
		respectively. Furthermore, although the $Z_{cs}(3985)$ and
		$Z_{cs}(4000)$ states exhibit a significant difference in width,
		these two resonances may originate from the same channel, the $S-$wave
		$(D_{s}\bar{D}^{*}+D_{s}^{*}\bar{D})/\sqrt{2}$. Additionally, we
		find two resonances in the $S-$wave $D_s^*\bar{D}^*$ channel, corresponding
		to the $Z_{cs}(4123)$ and $Z_{cs}(4220)$ states that await
		experimental confirmation.
		
	\end{abstract}
	\maketitle
	
	\section{Introduction}\label{sec:Introduction}
	In the past decade, ongoing experimental efforts have led to the
	discovery of a series of heavy quarkonium-like states known as the
	$XYZ$ states. The charged $Z$ states like $Z_c(3900)$ and
	$Z_c(4020)$ provide strong evidence of the exotic states, as they
	involve the light quarks to explain their non-zero electric charge.
	Experimental advancements in the $Z_b$ sector can be traced back to
	2011 when the Belle collaboration reported two charged exotic
	candidates, $Z_b(10610)$ and $Z_b(10650)$ \cite{Bondar2012}, which
	were later confirmed in subsequent studies
	\cite{Garmash2015,Garmash2016}. Multiple hidden-charm tetraquark
	candidates of the $Z_c$ states have been observed by the BESIII,
	Belle and CLEO collaborations in electron-positron annihilation,
	including the charged and neutral $Z_c(3900)$ and $Z_c(4020)$ states
	\cite{PhysRevLett.110.252001,Ablikim2013,Ablikim2014,Ablikim2014a,Ablikim2014b,Ablikim2015d,Ablikim2015,Ablikim2015b,Ablikim2015c,BelleCollaboration2013,Xiao2013}.
	These states, with their masses near the thresholds of
	$B^{(*)}\bar{B}^*$ and $D^{(*)}\bar{D}^*$, have been widely
	interpreted as the molecule states in the papers
	\cite{Molina2009d,Zhang2011,Sun2011,Ozpineci2013,He2013,Guo2013,Dong2013,Wang2014a,He2014,Aceti2014,Wang2014,Karliner2015,Baru2019,Wang2020,Ding2020,Dai2022}.
	Additionally, the existence of the strange partners with the $Q\bar
	Q s\bar q'$ ($q,q'=u,d$) configurations is predicted by the
	SU(3)-flavor symmetry, and indeed they have been discovered in
	recent years.
	
	In 2021, the BESIII collaboration observed an exotic hadron near the
	mass thresholds of $D_s^-D^{*0}$ and $D_s^{*-}D^{0}$ in the
	processes $e^+e^-\to K^+D_s^-D^{*0}$ and $K^+D_s^{*-}D^{0}$
	\cite{Ablikim2021}. The corresponding mass and width fitted with a
	Breit-Wigner line shape are
	\begin{eqnarray}
		&&M[Z_{cs}(3985)]=3982.2_{-2.6}^{+1.8}\pm2.1\ \text{MeV}\ \text{and}\nonumber\\
		&&\Gamma[Z_{cs}(3985)]=12.8_{-4.4}^{+5.3}\pm3.0 \ \text{MeV}.\label{BESIII 1}
	\end{eqnarray}
	Last year, they observed a neutral $Z_{cs}(3985)^0$ in the processes
	$e^+e^-\to K_S^0D_s^{+}D^{*-}$ and $K_S^0D_s^{*+}D^{-}$
	\cite{Ablikim2022}. The mass and width of the neutral
	$Z_{cs}(3985)^0$ have been determined to be ($3992.2\pm 1.7\pm1.6$)
	MeV and ($7.7^{+4.1}_{-3.8}\pm4.3$) MeV, respectively. Its mass,
	width and cross section are similar to those of the charged
	$Z_{cs}(3985)^+$, which suggests that the neutral $Z_{cs}(3985)^0$ is
	the isospin partner of the $Z_{cs}(3985)^+$. Furthermore, in 2021,
	the LHCb collaboration reported a series of distinct $Z_{cs}$
	states. In the hidden charm decay process $B^+\to J/\psi\phi K^+$,
	they observed two $Z_{cs}$ states with $J^P=1^+$ \cite{Aaij2021a}.
	One of these $Z_{cs}$ states is the $Z_{cs}(4000)^+$, which is
	discovered with high significance. Its mass and width are measured
	to be
	\begin{eqnarray}
		&&M[Z_{cs}(4000)]=4003\pm6_{-14}^{+4}\ \text{MeV}\ \text{and}\nonumber\\
		&&\Gamma[Z_{cs}(4000)]=131\pm15\pm26 \ \text{MeV},\label{LHCb 1}
	\end{eqnarray}
	respectively. Additionally, the other $Z_{cs}$ state,
	$Z_{cs}(4220)^+$, has a mass of $4216\pm24_{-30}^{+43}$ MeV and a
	width of $233\pm52_{-73}^{+97}$ MeV. The LHCb collaboration
	considers the $Z_{cs}(4000)^+$ and $Z_{cs}(3985)^+$ to be distinct
	states due to their apparently different widths, despite their close
	mass.
	
	This discovery of the exotic $Z_{cs}$ hadrons inspired various
	theoretical interpretations, including the compact tetraquark
	picture \cite{Wang2023,Maiani2021,Shi2021}, the molecule picture
	\cite{Chen2021,Yan2021,Meng2021,Zhai2022,Wu2022,Meng2022a,Du2022,Chen2022b},
	the mixing scheme
	\cite{Karliner2021a,Jin2021a,Yang2021,Han2022,Cao2023} and the cusp
	effect \cite{Luo2023}. When examining the BESIII and LHCb
	observations of the $Z_{cs}$ states, some authors of the Refs.
	\cite{Wu2022,Ortega2021a,Giron2021} proposed that the $Z_{cs}(3985)$
	and $Z_{cs}(4000)$ are the same entity, whereas the Refs.
	\cite{Chen2022b,Han2022,Maiani2021,Meng2021,Wang2023,Yang2021}
	considered them to be distinct hadrons. Moreover, one can gain further insights from the comprehensive reviews published in recent years \cite{Hosaka:2016pey,chenHiddencharmPentaquarkTetraquark2016,Chen:2016spr,Ali:2017jda,Esposito2017,Lebed:2016hpi,guoHadronicMolecules2018,liuPentaquarkTetraquarkStates2019,brambillaStatesExperimentalTheoretical2020,Lucha:2021mwx,dongSurveyHeavyHeavy2021,Brambilla:2021mpo,Meng:2022ozq,Chen:2022asf}.
	
	
	In Refs. \cite{Maiani2021,Meng2021}, the authors considered the
	$Z_{cs}(3985)$ and $Z_{cs}(4000)$ as the SU(3)-flavor partners of
	$Z_c(3900)$, whose neutral nonstrange members have opposite $C$
	parity. The authors suggested the $Z_{cs}(4000)/Z_{cs}(3985)$ is the
	pure molecular state composed of $(|\bar{D}_s^*D\rangle
	+/-|\bar{D}_sD^*\rangle)/\sqrt 2$. In addition, they also predicted
	the existence of a molecule composed of $\bar{D}_s^*D^*$ which may
	be confirmed by the BESIII in the subsequent experiment
	\cite{BESIIICollaboration2023}. However, the huge difference of
	their widths seems still hard to interpret.
	
	In this study, we employ the chiral effective field theory (ChEFT)
	to investigate the properties of the $Z_b$, $Z_c$ and $Z_{cs}$
	states in the molecular picture. To explore the existence and
	relationships of the possible resonances, we utilize the complex
	scaling method (CSM)
	\cite{aguilarClassAnalyticPerturbations1971,balslevSpectralPropertiesManybody1971b},
	which is a powerful tool that provides a consistent treatment of the
	bound states and resonances. We focus solely on the $S-$wave open-charm interaction, while neglecting the possible contributions from the
	hidden charm. As illustrated in our previous works
	\cite{Cheng2022b,Lin2022}, we consider the cross diagram
	$D\bar{D}^{*}\leftrightarrow D^{*}\bar{D}$ of the one-pion-exchange
	(OPE) contribution. This contribution introduces a complex potential
	arising from the three-body decay effect, which we take into account
	when investigating the widths of the resonances.
	
	This paper is organized as follows. In Sec. \ref{sec:framework}, we
	introduce our framework explicitly. In Sec. \ref{sec: potentials},
	we present the effective Lagrangians and potentials. In Sec.
	\ref{sec: results}, we solve the complex scaled Schr\"odinger
	equation and give the results of the $Z_b$, $Z_c$ and $Z_{cs}$. The
	last section \ref{sec:summary} is a brief summary.
	
	\section{Framework}\label{sec:framework}
	
	In this study, we consider the $Z_b$, $Z_c$ and $Z_{cs}$ states as
	the molecular systems with the quantum numbers
	$I^G(J^{PC})=1^+(1^{+-})$, $I^G(J^{PC})=1^+(1^{+-})$ and
	$I(J^{P})=1/2(1^{+})$, respectively. The specific molecule systems
	under investigation are $(B\bar{B}^*+B^{*}\bar{B})/\sqrt 2$,
	$B^*\bar{B}^*$, $(D\bar{D}^*+D^{*}\bar{D})/\sqrt 2$, $D^*\bar{D}^*$,
	$(D_s\bar{D}^*+D_s^{*}\bar{D})/\sqrt 2$ and $D_s^*\bar{D}^*$.
	
	In the earlier work \cite{Sun2011}, the $Z_b$ states were proposed
	as the bound states of
	$\left[B\bar{B}^{*}+B^*\bar{B}\right]/\sqrt{2}$ and $B^*\bar{B}^*$.
	The authors considered the $D$-wave channel and found that the $S-D$
	wave mixing effect could contribute significantly. Recent
	experiments \cite{Garmash2015,Garmash2016} have uncovered additional
	evidence supporting the interpretation of the $Z_b$ states as the
	resonances. These findings show that the masses of the $Z_b$ states
	are higher than the threshold of the $B^{(*)}\bar{B}^*$ pairs, and
	they can decay into the $B^{(*)}\bar{B}^*$ channel with the partial
	widths in the range of tens of MeV. These findings strongly favor
	the resonance interpretation over the bound state scenario.
	
	The present CSM work confirms that the $D$-wave channel has a
	minimal impact on the mass and width of the states. Therefore, we
	ignore the $D$-wave channel in this work. On the other hand, we find that
	the coupled channel effect between $(B\bar{B}^{*}+B^*\bar{B})/\sqrt
	2$ and $B^*\bar{B}^*$ is negligible for the near threshold states.
	In addition, there are inelastic channels in the final decay process, like $\Upsilon(nS)\pi$,
    that could be the constituents of the $Z_b$ states as well.
	However, the couplings strength between
	the $Z_b$ and the hidden-bottom channels is apparently smaller than that between $Z_b$ and the open-bottom channels.
	Therefore, the influence of the correction from the hidden-bottom channels should not be significant.
	Furthermore, for the $Z_c$ and $Z_{cs}$ systems, we adopt the same assumption that the inelastic hidden-heavy channels are not the primary constituents.
	As a result, we focus on the simplest case, considering only the $S-$wave open-heavy single channel.

	The masses of the charmed meson and exchanged light mesons are
	collected in Table \ref{tab: mass meson}. We take the isospin
	average masses to deal with the isospin conserving process.

	\begin{table}[htbp]
		\begin{tabular}{cccc}\hline\hline
			Mesons&Mass(MeV)&Mesons&Mass(MeV)\\
			\hline
			$D^{+}$&1869.66&$B^{*}$&5324.70\\
			$D^{0}$&1864.84&$D_{s}^+$&1968.34\\
			$D^{*+}$&2010.26&$D_{s}^{*+}$&2112.2\\
			$D^{*0}$&2006.85&$\pi^{\pm}$&139.57\\
			$B^{+}$&5279.34&$\pi^{0}$&134.98\\
			$B^{0}$&5279.65&&\\
			\hline\hline
		\end{tabular}
		\caption{The masses of the charmed, bottomed and pion mesons, which are taken from Ref. \cite{ParticleDataGroup2020}.}\label{tab: mass meson}
	\end{table}
	
	
	\subsection{A brief discussion on the CSM}\label{subsec:csm}
	
	We first provide a brief overview of the CSM proposed by Aguilar,
	Balslev and Combes in the 1970s
	\cite{aguilarClassAnalyticPerturbations1971,balslevSpectralPropertiesManybody1971b},
	commonly known as the ABC theorem. The CSM is a powerful approach
	that allows for the treatment of resonances in a manner similar to
	the bound states. The transformation of the radial coordinate $r$
	and its conjugate momentum $k$ in the CSM are defined by:
	\begin{eqnarray}
		U(\theta)r=re^{i\theta},\qquad U(\theta)k=ke^{-i\theta}. \label{eq:rktrans}
	\end{eqnarray}
	After the complex scaling operation, the Schr\"odinger equation
	\begin{eqnarray}
		\frac{p^2}{2m}\phi_l(p)+ \int \frac{p'^{2}dp'}{(2\pi)^3} V_{l,l'} (p,p')\phi_{l'}(p')=E\phi_l(p) \label{MSE}
	\end{eqnarray}
	in the momentum space becomes
	\begin{eqnarray}
		&& \frac{p^2e^{-2i\theta}}{2m}\tilde{\phi}_l(p)+ \int \frac{p'^{2}e^{-3i\theta}dp'}{(2\pi)^3} V_{l,l'} (pe^{-i\theta},p'e^{-i\theta})\tilde{\phi}_{l'}(p') \nonumber\\
		&&=E\tilde{\phi}_l(p), \label{eq:SECSM}
	\end{eqnarray}
	with the normalization relation
	\begin{eqnarray}
		&&  \frac{e^{-3i\theta}}{(2\pi)^3}\int_{0}^{\infty}\tilde{\phi}_l(p)^2 p^2dp=1,\label{eq:NM}
	\end{eqnarray}
	where $l,l'$ are the orbital angular momenta, and $p$ represents the
	momentum in the center-of-mass frame. The potential $V_{l,l'}$ after
	partial wave decomposition can be expressed as
	\begin{eqnarray}
		V_{l,l'}&=& \int d\boldsymbol{\Omega}'\int d\boldsymbol{\Omega}\sum_{m_{l'}=-l'}^{l'}\langle l',m_{l'};s,m_j-m_{l'}|j,m_j\rangle\nonumber\\
		&\times&\sum_{m_{l}=-l}^{l}\langle l,m_{l};s,m_j-m_{l}|j,m_j\rangle \mathcal{Y}_{l',m_{l'}}^*(\theta',\phi') \nonumber\\
		&\times& \mathcal{Y}_{l,m_{l}}(\theta,\phi)\langle s,m_j-m_{l'}|\mathcal{V}|s,m_j-m_l\rangle,\label{eq:Vpartial}
	\end{eqnarray}
	where $s$ and $j$ represent the total spin and total angular
	momentum of systems, $m_l$ is the corresponding magnetic quantum
	number. The $\mathcal{Y}_{l,m_{l}}(\theta,\phi)$ represents the
	spherical harmonics associated with the angular coordinates
	$\theta$, $\phi$. The potential operator $\mathcal{V}$ acts on the
	states $|s,m_j-m_{l'}\rangle$ and $|s,m_j-m_l\rangle$.
	
\begin{figure}[htbp]
		\includegraphics[width=210pt]{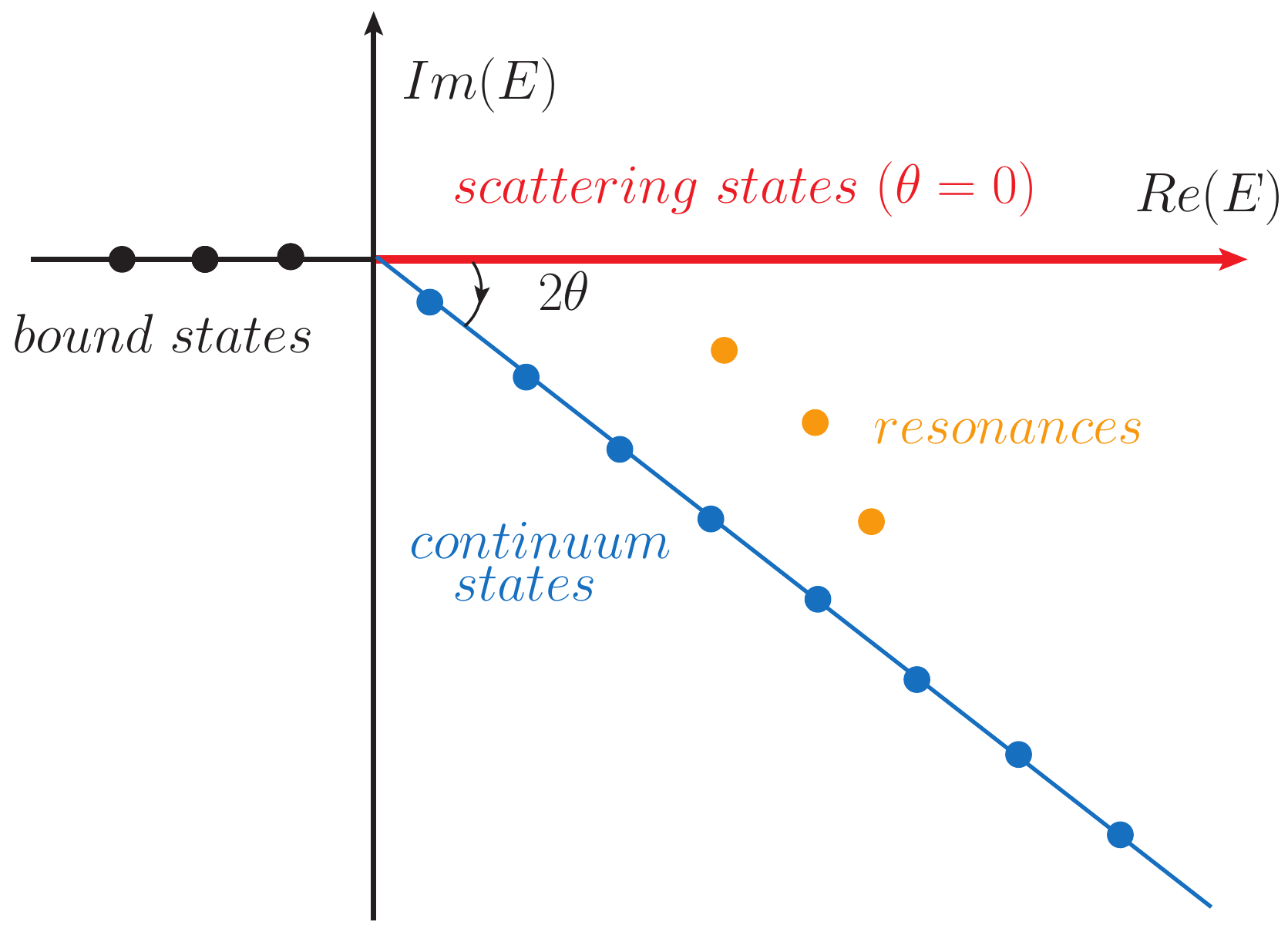}
		\caption{The eigenvalue distribution of the complex scaled Schr\"odinger equation for the two-body systems. }\label{fig: CSM plot}
	\end{figure}

	After performing the complex scaling operation, the resonance pole
	crosses the branch cut into the first Riemann sheet when the
	rotation angle $\theta$ reaches a sufficiently large value, as
	depicted in Fig. \ref{fig: CSM plot}. Consequently, the wave
	functions of the resonances become square-integrable,  similar to
	those of the normalizable bound states. Further information on this
	technique can be found in Refs.
	\cite{aoyamaComplexScalingMethod2006,hoMethodComplexCoordinate1983}.
	
	\subsection{Analyticity of the OPE potentials for the $DD^*$ system}\label{subsec:anlyticity Hcc}

	In our previous works \cite{Cheng2022b, Lin2022}, we investigated
	the double-charm tetraquark system using the CSM method. Notably, we
	found that the $D\bar{D}^*$ system exhibits a unique characteristic where
	the zeroth component of the transferred momentum of the exchanged
	pion exceeds the pion mass. This leads to an imaginary part in the
	OPE potential. If a pole is obtained in this system, it would
	correspond to an energy with an imaginary part, which can be
	interpreted as its half-width. In the current study, we encounter
	this situation when examining the OPE potential in the
	$(D\bar{D}^*+D^{*}\bar{D})/\sqrt 2$ system with $1^+(1^{+-})$.
	
	When considering the process $D\bar{D}^*\to D^* \bar D$, one can get
	the OPE potential as follows
	\begin{eqnarray}
		V_\pi\propto\frac{g^2}{2f_\pi^2}\frac{(\boldsymbol{\epsilon}^*\cdot\boldsymbol{q})(\boldsymbol{\epsilon}\cdot\boldsymbol{q})}{\boldsymbol{q}^2+m_\pi^2-q_0^2}, \label{eq:OPE}
	\end{eqnarray}
	where $q$ represents the transferred momentum of the pion, and $q_0$
	is its zeroth component. Since $q_0 \approx m_{D^*} - m_{D} >
	m_\pi$, the poles of the OPE potential are located on the real
	transferred momentum axis. However, when performing the integral
	along the real $p'$ axis in Eq. \eqref{MSE}, we encounter a
	numerical divergence. Fortunately, the CSM can resolve this
	divergence issue without altering the analyticity of the OPE
	potential. Through a complex scaling operation, the pole of the OPE
	potential is rotated away from the real momentum axis in the
	momentum plane. As a result, the integral along the real momentum
	axis bypasses the pole, effectively avoiding divergence.
	
\begin{figure}[htbp]
		\includegraphics[width=220pt]{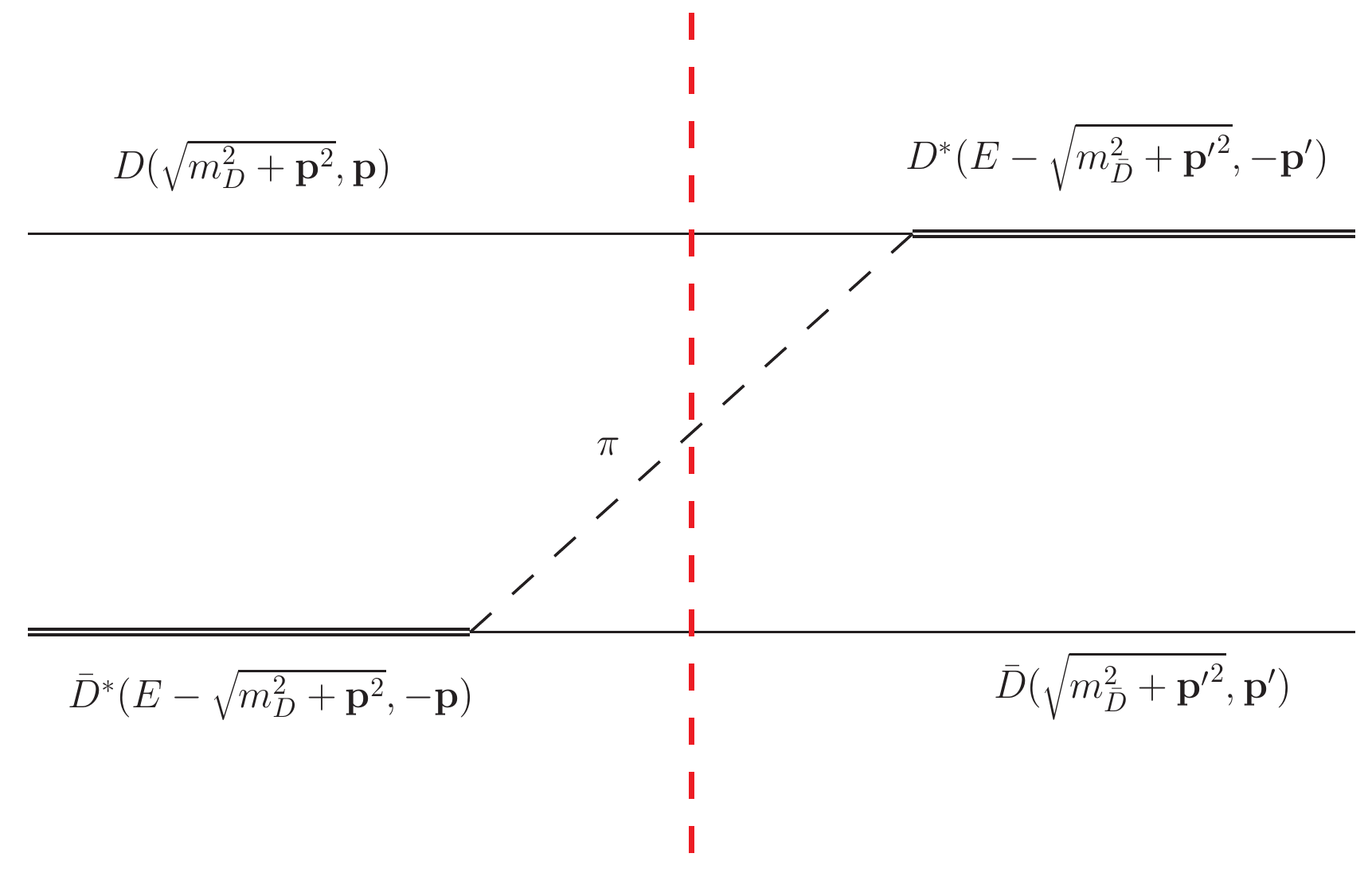}
		\caption{Three-body intermediate diagram in the process $D\bar{D}^{*}\to D^*\bar{D}$. The total energy of the $D\bar{D}^{*}/D^*\bar{D}$ is $E$, and the mesons which are cut by the red dashed line are on shell.}\label{fig: FD Tcc}
	\end{figure}

	As shown in Fig. \ref{fig: FD Tcc}, we denote the total energy of
	the $D\bar{D}^*/D^*\bar{D}$ system as $E$ and assume the $D$ meson
	to be on-shell. In this case, the expression for $q_0$ is given by
	$q_0=E-\sqrt{m_D^2+\boldsymbol{p}^2}-\sqrt{m_{\bar{D}}^2+\boldsymbol{p'}^2}$.
	With the heavy quark approximation, we neglect the kinetic energy
	contribution to $q_0$ and introduce an energy shift $E\to
	E+m_D+m_{D^*}$. As a result, we obtain $q_0=E+m_{D^*}-m_D$.

	In other processes, the three-body effect vanishes, and we should
	consider the different values of $q_0$ in the OPE potential. The
	specific values of $q_0$ for each case are summarized in Table
	\ref{tab: q0}.
	
	\begin{table}[htbp]
		\begin{tabular}{cccc}\hline\hline
			Process&$D\bar{D}^*\to D^*\bar{D}$&$D^*\bar{D}^*\to D^*\bar{D}^*$&$B\bar{B}^*\to B^*\bar{B}$\\
			$q_0$&$E+m_{D^*}-m_{D}$&$0$&$m_{B^*}-m_{B}$\\
			\hline\hline
		\end{tabular}
		\caption{The $q_0$ is the zeroth component of the transferred
			momentum. $E$ is the total energy relative to the corresponding
			threshold. The other cases not listed all give $q_0=0$.}\label{tab:
			q0}
	\end{table}

	\section{Lagrangians and Potentials}\label{sec: potentials}
	
	For the interaction of two heavy mesons, the chiral effective
	Lagrangians are constructed based on the heavy quark symmetry and
	SU(3)-flavor symmetry. The explicit expressions are given by {\small
		\begin{eqnarray}
			\mathcal{L} & = &-i\langle H^{(Q)}_bv\cdot(
			\delta_{ba}\partial+i\Gamma_{ba})\bar{H}^{(Q)}_a\rangle+g\langle H^{(Q)}_b\mathbb{A}_{ba}^\mu\gamma_\mu\gamma_5\bar{H}_a^{(Q)}\rangle\nonumber \\
			&& -i\langle\bar{\tilde{H}}_b^{(\bar Q)}v\cdot(
			\delta_{ba}\partial+i\Gamma_{ba})\tilde{H}_a^{(\bar Q)}\rangle+g\langle\bar{\tilde{H}}_b^{(\bar Q)}\mathbb{A}_{ba}^\mu\gamma_\mu\gamma_5\tilde{H}_a^{(\bar Q)}\rangle\nonumber\\
	\end{eqnarray}}
	where $H^{(Q)}$ is defined as
	\begin{eqnarray}
		H_a^{(Q)} &=& \frac{1+\slashed{v}}{2}\left[P_a^{*\mu}\gamma_{\mu}-P_a\gamma_5\right].
	\end{eqnarray}
	And $\bar{H}_a^{(Q)}$, $\tilde{H}^{(\bar Q)}$ and
	$\bar{\tilde{H}}_a^{(\bar Q)}$ are
	\begin{eqnarray}
		\bar{H}_a^{(Q)} &=&\gamma_0 H_a^{(Q)\dagger}\gamma_0=\left[P_a^{*\dagger\mu}\gamma_{\mu}+P_a^\dagger\gamma_5\right]\frac{1+\slashed{v}}{2},\nonumber\\
		\tilde{H}^{(\bar Q)}&=&\left[\tilde{P}_a^{*\mu}\gamma_\mu+\tilde{P}_a\gamma_5\right]\frac{1-\slashed{v}}{2}\quad \text{and}\nonumber\\
		\bar{\tilde{H}}_a^{(\bar Q)}&=&\gamma_0\tilde{H}^{(\bar Q)\dagger}\gamma_0=\frac{1-\slashed{v}}{2}\left[\tilde{P}_a^{*\dagger\mu}\gamma_\mu-\tilde{P}_a^\dagger\gamma_5\right],
	\end{eqnarray}
	respectively, with $P_a^{(*)}=\left(D^{(*)0},D^{(*)+},D_s^{(*)+}\right)$ and $\tilde{P}_a^{(*)}=\left(D^{(*)-},\bar{D}^{(*)0},\bar{D}_s^{(*)-}\right)$.
	
	The light meson concerned parts are given that
	\begin{eqnarray}
		\mathbb{A}_\mu & = &\frac{i}{2} [\xi^\dagger(\partial_\mu\xi)+(\partial_\mu\xi)\xi^\dagger],\quad \Gamma_\mu=\frac{i}{2} [\xi^\dagger(\partial_\mu\xi)-(\partial_\mu\xi)\xi^\dagger],\nonumber\\
		\xi  &=& \text{exp}[\frac{i\mathcal{M}}{f_\pi}]\quad \text{and} \\
		\mathcal{M} & = & \begin{pmatrix}
			\frac{\pi^0}{\sqrt{2}}+\frac{\eta}{\sqrt{6}} & \pi^+ & K^+ \\
			\pi^- & \frac{\pi^0}{\sqrt{2}}+\frac{\eta}{\sqrt{6}} & K^0 \\
			K^- & \bar{K}^0 & -\frac{2}{\sqrt{6}}\eta
		\end{pmatrix},
	\end{eqnarray}
	where the pion decay constant $f_\pi$ is equal to 132 MeV. The
	coupling constant associated with the $\pi$ exchange is $g=0.59$
	\cite{ahmedFirstMeasurement2001}.

	The corresponding OPE potential in momentum space can be expressed
	as follows
	\begin{eqnarray}
		V^{{D\bar{D}^*}/{B\bar{B}^*}}&=&-\frac{g^2}{2f_\pi^2}\frac{(\boldsymbol{\epsilon}^*\cdot\boldsymbol{q})(\boldsymbol{\epsilon}\cdot\boldsymbol{q})}{\boldsymbol{q}^2+m_\pi^2-q_0^2},\nonumber\\
		V^{D^*\bar{D}^*/B^*\bar{B}^*}&=&-\frac{g^2}{2f_\pi^2}\frac{(\boldsymbol{\mathcal{T}}_1\cdot\boldsymbol{q})(\boldsymbol{\mathcal{T}}_2\cdot\boldsymbol{q})}{\boldsymbol{q}^2+m_\pi^2-q_0^2},\nonumber\\
	\end{eqnarray}
	where $\boldsymbol{\mathcal{T}}_1$ and $\boldsymbol{\mathcal{T}}_2$
	represent the spin 1 operator with the forms
	$\boldsymbol{\mathcal{T}}_1=-i\boldsymbol{\epsilon}_3^\dagger\times\boldsymbol{\epsilon}_1$
	and
	$\boldsymbol{\mathcal{T}}_2=-i\boldsymbol{\epsilon}_4^\dagger\times\boldsymbol{\epsilon}_2$.
	Since we focus solely on the $S$-wave interactions, we can replace
	the above spin-dependent operator with
	$(\boldsymbol{\epsilon}^*\cdot
	\boldsymbol{q})(\boldsymbol{\epsilon}\cdot
	\boldsymbol{q})\to\frac{1}{3}\boldsymbol{q}^2$ and
	$(\boldsymbol{T}_1\cdot \boldsymbol{q})(\boldsymbol{T}_2\cdot
	\boldsymbol{q})\to\frac{1}{3}\boldsymbol{q}^2\boldsymbol{T}_1\cdot\boldsymbol{T}_2$.

	Regarding the contact term interaction, we adopt the form derived in
	Ref. \cite{Wang2020}. Upon performing the partial wave
	decomposition, one can obtain the $S$-wave contact potential as
	\begin{eqnarray}
		\left[V_{ct}\right]_{l,l'}&=&\tilde{C}_s+C_s(p^2+p'^2),\nonumber
	\end{eqnarray}
	where $\tilde{C}_s$ and $C_s$ represent the partial wave low energy
	constants (LECs). We restrict our analysis to the lowest-order
	interaction and do not consider higher-order effects, such as the
	one-loop contribution.
	
	To obtain the effective potentials, we introduce a Gaussian
	regulator to the potentials as follows
	\begin{eqnarray}
		V_{l,l'}=V_{l,l'}\exp\left(-\frac{p'^2}{\Lambda^2}-\frac{p^2}{\Lambda^2}\right),
	\end{eqnarray}
	where $\Lambda$ is the cutoff parameter. The parameters $\Lambda$, $\tilde{C}_s$ and $C_s$ can be adjusted while keeping the coupling constants in the OPE potential fixed.

	\section{Numerical results}\label{sec: results}
	
	During the numerical calculation process, we discretize the
	Schr\"odinger Eq. \eqref{MSE} in momentum space using the Gaussian
	quadrature approach. We approximate the integral over the potential
	as a weighted sum over $N$ integration points for $p=k_j$ ($j=1,N$):
	\begin{eqnarray}
		&&\int_0^{\infty}dp'p'^{2}V(p,p')\phi(p')\simeq\sum_{j=1}^{N}\omega_j p_j^2V(p,p_j)\phi(p_j), \nonumber\\
		&&\frac{p^2}{2m}\phi(p)+\frac{1}{(2\pi)^3}\sum_{j=1}^{N}\omega_j p_j^2V(p,p_j)\phi(p_j)=E\phi(p),\nonumber\\\label{discretize}
	\end{eqnarray}
	where $p_j$ and $\omega_j$ represent the Gaussian quadrature points
	and weights, respectively. Furthermore, for clarity, we will omit
	the orbital angular momentum subscript from this point onward. In
	Eq. \eqref{discretize}, we have $N$ unknowns $\phi(k_j)$ and an
	unknown $\phi(k)$. To avoid the need to determine the entire
	function $\phi(k)$, we restrict the solution to the same values of
	$k_i$ used to approximate the integral. This lead to $N$ coupled
	linear equations:
	\begin{eqnarray}
		&&\frac{p_i^2}{2m}\phi(p_i)+\frac{1}{(2\pi)^3}\sum_{j=1}^{N}\omega_j p_j^2V(p_i,p_j)\phi(p_j)=E\phi(p_i).\nonumber\\\label{discretize N2}
	\end{eqnarray}
	Therefore, the Schr\"odinger equation can be expressed in matrix
	form as
	\begin{eqnarray}
		&&[H][\phi]=E[\phi],
	\end{eqnarray}
	with explicit matrices form
	\begin{widetext}
		\begin{eqnarray}
			&&\begin{pmatrix}
				\frac{p_1^2}{2m}+\frac{1}{(2\pi)^3}\omega_1 p_1^2V(p_1,p_1) & \frac{1}{(2\pi)^3}\omega_2 p_2^2V(p_1,p_2) & \cdots  & \frac{1}{(2\pi)^3}\omega_N p_N^2V(p_1,p_N) \\
				\frac{1}{(2\pi)^3}\omega_1 p_1^2V(p_2,p_1) & \frac{p_2^2}{2m}+\frac{1}{(2\pi)^3}\omega_2 p_2^2V(p_2,p_2) & \cdots &  \\
				\vdots &  & &\\
				\frac{1}{(2\pi)^3}\omega_1 p_1^2V(p_N,p_1)&\cdots&\cdots&\frac{p_N^2}{2m}+\frac{1}{(2\pi)^3}\omega_N p_N^2V(p_N,p_N)
			\end{pmatrix}     \begin{pmatrix}
				\phi(p_1)\\
				\phi(p_2)\\
				\vdots\\
				\phi(p_N)\\
			\end{pmatrix} =  E\begin{pmatrix}
				\phi(p_1)\\
				\phi(p_2)\\
				\vdots\\
				\phi(p_N)\\
			\end{pmatrix},\nonumber\\
		\end{eqnarray}
	\end{widetext}
	where the wave function $\phi(k)$ on the grid can be represented as
	the $N\times1$ vector $[\phi(p_i)]=\begin{pmatrix}
		\phi(p_1)&\phi(p_2)&\cdots&\phi(p_N)\\
	\end{pmatrix}^T$.
	Then, we can effectively solve Eq. \eqref{MSE}. To find solutions
	for the complex Schr\"odinger Eq. \eqref{eq:SECSM}, we can simply
	make the substitutions $p_i\to p_i e^{-i\theta}$,
	$\omega_i\to\omega_i e^{-i\theta}$ and
	$\phi(p_i)\to\tilde{\phi}(p_i)$.
	
	\subsection{The $Z_b$ and $Z_c$ system}\label{sec:Zb and Zc}

	\begin{figure}[htbp]
		\subfigure[]{ \label{fig: ur obep re}
			\includegraphics[width=180pt]{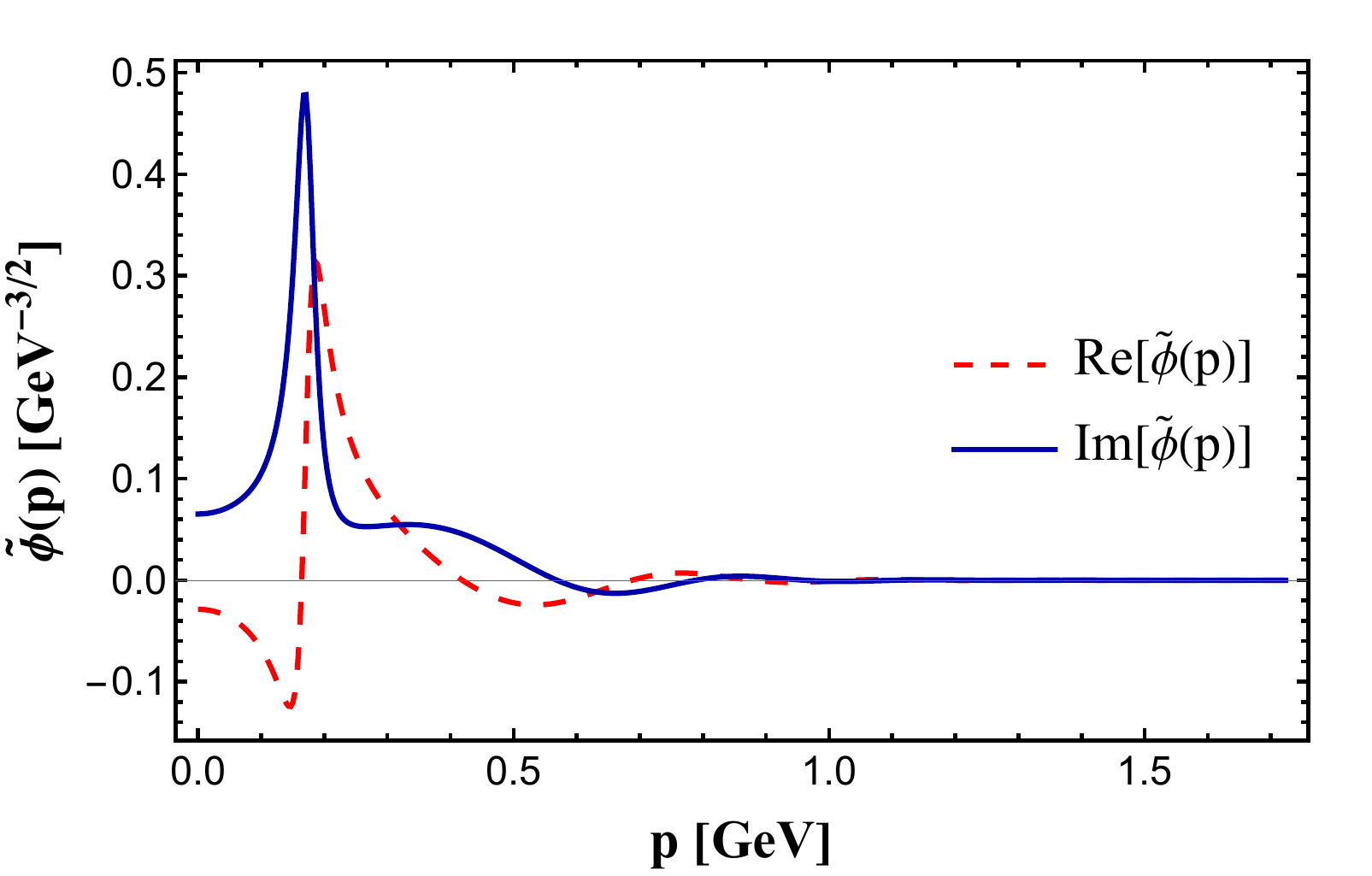}}\hspace{40pt}
		\subfigure[]{ \label{fig: ur obep im}
			\includegraphics[width=180pt]{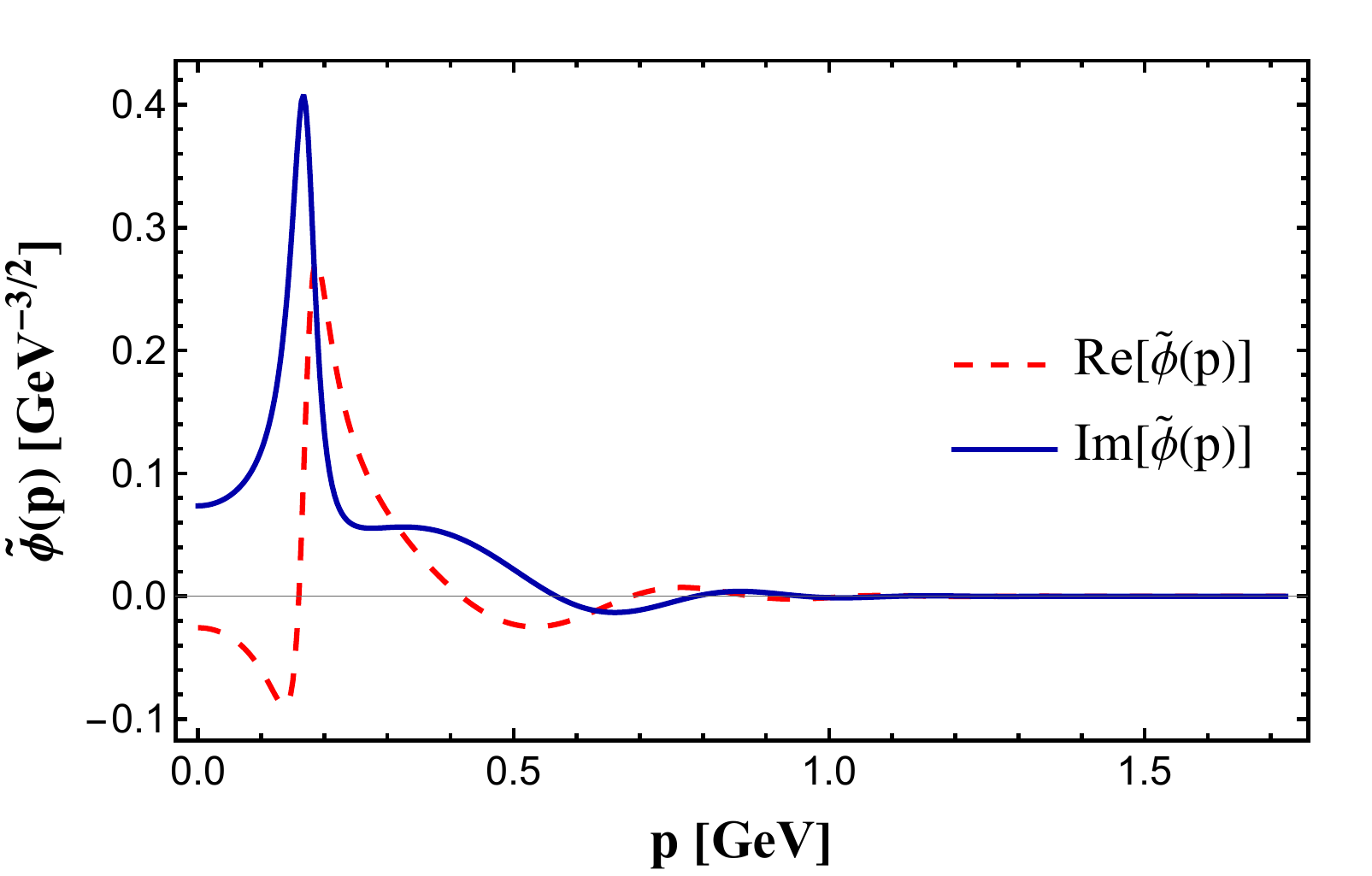}}
		\caption{\small The wave functions $\tilde{\phi}(p)$ of the
			$Z_c$ state with the $I^G(J^{PC})=1^+(1^{+-})$. The rotation angle
			$\theta=35^\circ$ and the parameters $\Lambda=0.300$ GeV, $\tilde{C}_{s}=2.86\times10^2$ GeV$^{-2}$ and $C_{s}=-59.9\times10^2$ GeV$^{-4}$. The two diagrams correspond to system (a) $\left[D\bar{D}^{*}+D^*\bar{D}\right]/\sqrt{2}$ and (b) $D^*\bar{D}^*$ respectively.}\label{fig: urobep}
	\end{figure}
	
	In this subsection, we investigate the exotic hadrons $Z_b$ and
	$Z_c$ using ChEFT. A similar study has been performed in Ref.
	\cite{Wang2020}, where the $Z_c(3900)$ and $Z_c(4020)$ ($Z_b(10510)$
	and $Z_b(10650)$) are interpreted as
	$\left[D\bar{D}^{*}+D^*\bar{D}\right]/\sqrt{2}$ and $D^*\bar{D}^*$
	($\left[B\bar{B}^{*}+B^*\bar{B}\right]/\sqrt{2}$ and $B^*\bar{B}^*$)
	molecule with $J^P=1^+(1^{+-})$, respectively. However, in our
	present work using CSM, we find that the contributions from the
	$D$-wave constituents are negligible. Therefore, we neglect the
	$S-D$ mixing effect and solely focus on the $S$-wave channel in this
	section.
	
	In our analysis, as shown in Table \ref{tab: results}, we perform a
	fit of the LECs for the two $Z_b$ and $Z_c$ states. Comparing our
	results with those in Ref. \cite{Wang2020}, we find a similar cutoff
	value $\Lambda$ within a reasonable range. However, the LECs
	$\tilde{C}_s$ and $C_s$ exhibit some variations, which could be
	attributed to our omissions of the $D$-wave channel and the higher
	order contribution. Additionally, we calculate the root-mean-square
	(RMS) radii, as shown in Table \ref{tab: results}, and find that the
	sizes of the $Z_b$ states are smaller than those of the $Z_c$
	states. Interestingly, the sizes of the two $Z_b$ ($Z_c$) states are
	nearly identical. Moreover, the corresponding wave functions, as
	depicted in Fig. \ref{fig: urobep}, exhibit a striking resemblance.
	This phenomenon is reasonable since our analysis in this work does
	not account for the higher-order spin-dependent correction terms.
	The satisfaction of the heavy quark spin symmetry justifies the
	similarities in the energy, decay width, size and wave function
	observed in the $Z_b$ and $Z_c$ states.
	
	As discussed in Ref. \cite{Cheng2022b}, the $DD^*/D\bar{D}^*$ system
	considered as the $T_{cc}^{+}/X(3872)$ state can decay into the
	three-body open-charm channels $DD\pi/D\bar{D}\pi$. In the case of the
	isovector $D\bar{D}^*$ system, it is also necessary to consider the
	influence of the three-body decay. The numerical results in the
	scheme we adopt, shown in Table \ref{tab: results} (row "Adopt"),
	are very close to the results under the instantaneous approximation
	$q_0=0$. This implies that the mass, width and size have minimal
	changes. The reason why the choice of $q_0$ matters for the
	$T_{cc}^{+}$ system but not for the $Z_c$ system can be understood
	as follows. The mass of the $T_{cc}^{+}$ state is below the
	threshold of the $DD^*$ system, making the two-body decay process
	kinetically forbidden. Therefore, the three-body decay becomes the
	dominant decay modes, and the value of $q_0$, which partly reflects
	the three-body decay width, becomes important. On the other hand,
	the $Z_c(3900)^+$ state is clearly above the threshold of the
	$D\bar{D}^*$ system, allowing for the two-body decay process. Since
	the contribution from the three-body decay is significantly smaller
	in this case, the choice of $q_0$ does not significantly alter the
	results.

			\begin{widetext}
		\begin{center}
			\begin{table}[htbp]
				\renewcommand{\arraystretch}{1.2}{
					\setlength\tabcolsep{4pt}{
						\begin{tabular}{ccccc}\hline\hline
							\centering
							System&Threshold&$\left[m,\Gamma\right]_{\text{pole}}$(MeV)&$\left[m,\Gamma\right]_{\text{exp}}$(MeV)&RMS(fm)\\\hline
							$\left[B\bar{B}^{*}+B^*\bar{B}\right]/\sqrt{2}$&10604.2&$\left[10606.9_{-1.5}^{+1.8},15.0_{-3.2}^{+3.4}\right]$&$\left[10607.2_{-2.0}^{+2.0},18.4_{-2.4}^{+2.4}\right]$&$0.70_{-0.01}^{+0.07}-0.15_{-0.10}^{+0.09}i$\\
							$B^*\bar{B}^*$&10649.4&$\left[10652.2_{-1.6}^{+1.8},14.8_{-3.2}^{+3.4}\right]$&$\left[10652.2_{-1.5}^{+1.5},11.5_{-2.2}^{+2.2}\right]$&$0.70_{-0.02}^{+0.07}-0.15_{-0.11}^{+0.09}i$\\
							$\left[D\bar{D}^{*}+D^*\bar{D}\right]/\sqrt{2}$(Adopt)&3875.8&$\left[3884.3_{-0.6}^{+0.6},26.0_{-1.4}^{+1.4}\right]$&$\left[3881.7_{-2.3}^{+2.3},26.6_{-3.4}^{+3.0}\right]$&$1.21_{-0.05}^{+0.06}+0.12_{-0.03}^{+0.03}i$\\
							$\left[D\bar{D}^{*}+D^*\bar{D}\right]/\sqrt{2}$(Inst)&3875.8&$\left[3884.8_{-0.6}^{+0.6},25.8_{-1.4}^{+1.4}\right]$&$\left[3881.7_{-2.3}^{+2.3},26.6_{-3.4}^{+3.0}\right]$&$1.20_{-0.05}^{+0.06}+0.13_{-0.03}^{+0.03}i$\\
							$D^*\bar{D}^*$&4017.1&$\left[4025.8_{-0.6}^{+0.6},24.0_{-1.4}^{+1.3}\right]$&$\left[4025.5_{-5.6}^{+3.7},26.0_{-6.0}^{+6.0}\right]$&$1.20_{-0.05}^{+0.06}+0.13_{-0.03}^{+0.03}i$\\
							\hline\hline
				\end{tabular}}}
				\caption{The extracted poles for all states are listed with the quantum numbers $I^G(J^{PC})=1^+(1^{+-})$. The fitted parameter for the $B^*\bar{B}^{(*)}$ system are $\Lambda=0.510^{+0.027}_{-0.041}$ GeV, $\tilde{C}_{s}=0.48^{+0.15}_{-0.13}\times10^2$ GeV$^{-2}$ and $C_{s}=-5.4^{+0.63}_{-0.65}\times10^2$ GeV$^{-4}$. The fitted parameter for the $D^*\bar{D}^{(*)}$ system are $\Lambda=0.300^{+0.012}_{-0.013}$ GeV, $\tilde{C}_{s}=2.86^{+0.21}_{-0.22}\times10^2$ GeV$^{-2}$ and $C_{s}=-59.9^{+2.8}_{-3.1}\times10^2$ GeV$^{-4}$. The RMS is the root-mean-square radius in the CSM, which has been discussed in the Ref. \cite{hommaMatrixElementsPhysical1997}. Its real part is interpreted as an expectation value, and the imaginary part corresponds to a measure of the uncertainty in observation. The data of row Adopt are the results we actually adopt, the $q_0$ herein is from Table \ref{tab: q0}. The data of row Inst are from the instantaneous approximation $q_0=0$.}	\label{tab: results}
			\end{table}
		\end{center}
	\end{widetext}

	\begin{widetext}
	\begin{center}
		\begin{table}[htbp]
			\renewcommand{\arraystretch}{1.2}{
				\setlength\tabcolsep{4pt}{
					\begin{tabular}{ccccc}\hline\hline
						\centering
						System&Threshold&$\left[m,\Gamma\right]_{\text{pole}}$(MeV)&$\left[m,\Gamma\right]_{\text{exp}}$(MeV)&RMS(fm)\\\hline
						$\left[D_s\bar{D}^{*}+D_s^*\bar{D}\right]/\sqrt{2}$(1*)&3976.1&$\left[3982.4^{+2.2}_{-2.1},14.1^{+3.7}_{-3.6}\right]$&$\left[3982.5_{-3.3}^{+2.8},12.8_{-5.3}^{+6.1}\right]$&$1.89^{+0.13}_{-0.11}+0.43^{+0.09}_{-0.14}i$\\
						$\left[D_s\bar{D}^{*}+D_s^*\bar{D}\right]/\sqrt{2}$(2*)&&$\left[4010.7^{+6.3}_{-6.2},119.6^{+14.5}_{-14.7}\right]$&$\left[4003_{-15.2}^{+7.2},131_{-30.0}^{+30.0}\right]$&$1.78^{+0.18}_{-0.14}+1.31^{+0.08}_{-0.07}i$\\
						$D_s^*\bar{D}^{*}$(1)&4119.1&$\left[4125.2_{-2.1}^{+2.2},13.2_{-3.4}^{+3.5}\right]$&$\left[4123.5_{-1.3}^{+1.3},-\right]$&$1.89_{-0.11}^{+0.13}+0.43_{-0.14}^{+0.09}i$\\
						$D_s^*\bar{D}^{*}$(2)&&$\left[4152.7_{-6.0}^{+6.1},115.0_{-14.2}^{+14.0}\right]$&$\left[4216_{-38}^{+49},233_{-90}^{+110}\right]$&$1.78_{-0.14}^{+0.18}+1.31_{-0.07}^{+0.08}i$\\
						\hline\hline
			\end{tabular}}}
			\caption{The poles are all listed with the number $I(J^{P})=1/2(1^{+})$. The fitted parameter for the $D_s^*\bar{D}^{(*)}$ system are $\Lambda=0.192^{+0.012}_{-0.013}$ GeV, $\tilde{C}_{s}=6.8^{+2.8}_{-2.7}\times10^2$ GeV$^{-2}$ and $C_{s}=-186.9^{+50.4}_{-64.4}\times10^2$ GeV$^{-4}$. The RMS is the root-mean-square radius in the CSM, which has been discussed in the Ref. \cite{hommaMatrixElementsPhysical1997}. Its real part is interpreted as an expectation value, and the imaginary part corresponds to a measure of the uncertainty in observation. The states labeled as ``1*'' and ``2*'' correspond to the input states. The symbol ``-" indicates that the width of the $Z_{cs}(4123)$ state has not been confirmed by experiment yet.}\label{tab: results2}
		\end{table}
	\end{center}
\end{widetext}

	\subsection{The $Z_{cs}$ system}\label{{sec:Zcs}}
	
			\begin{figure}[htbp]
		\subfigure[]{ \label{fig: Zcs Dis 1}
			\includegraphics[width=250pt]{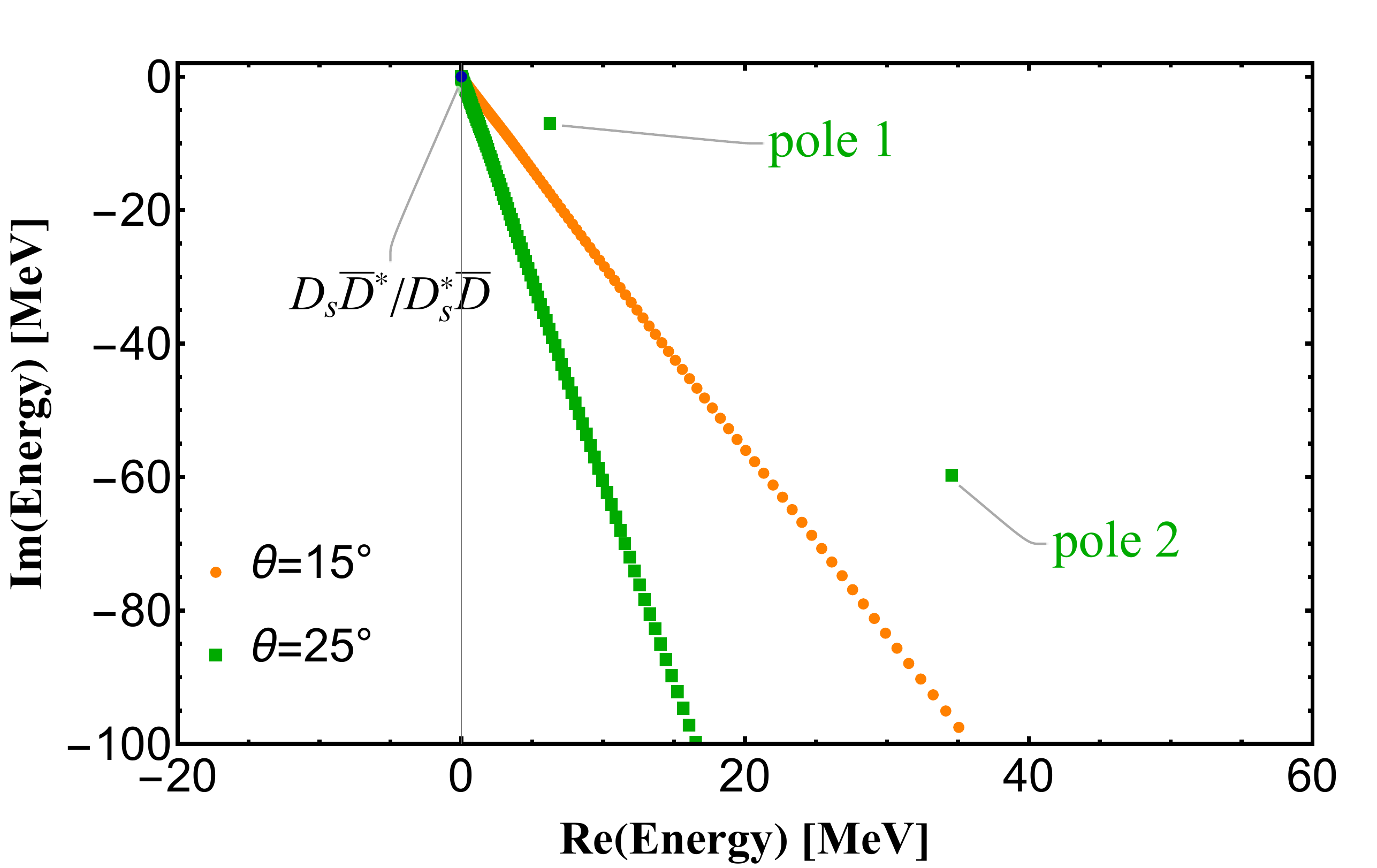}}\hspace{40pt}
		\subfigure[]{ \label{fig: Zcs Dis 2}
			\includegraphics[width=250pt]{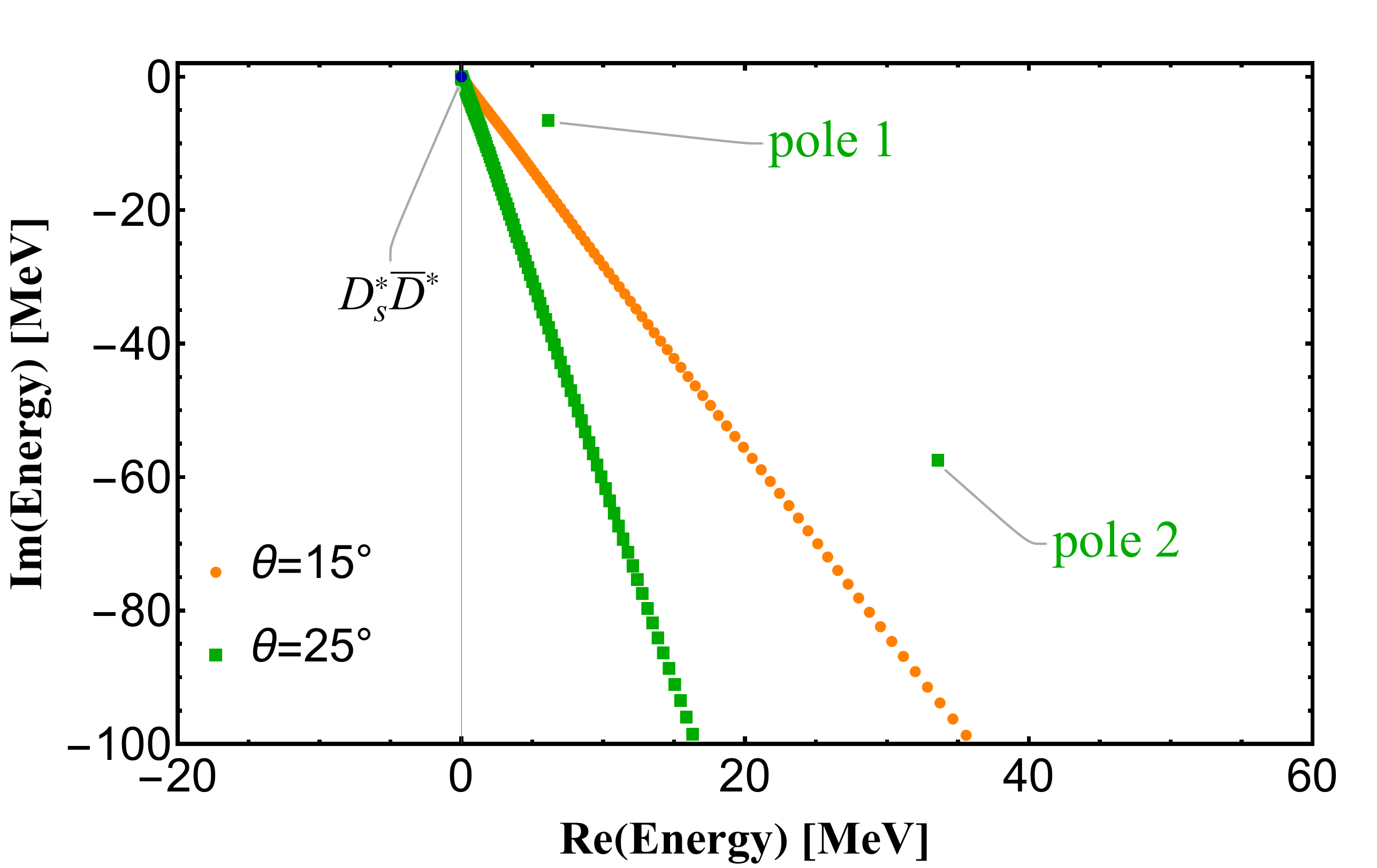}}
		\caption{\small The eigenvalue distribution of the $Z_{cs}$
			with the $I(J^P)=1/2(1^+)$.
			The parameters $\Lambda=0.192$ GeV, $\tilde{C}_{s}=6.8\times10^2$ GeV$^{-2}$ and $C_{s}=-186.9\times10^2$ GeV$^{-4}$.The orange (green) points (square points) and lines correspond to the complex rotation angle
			$\theta=35^\circ$ ($40^\circ$). The two diagrams correspond to system: (a) $\left[D_s\bar{D}^{*}+D_s^*\bar{D}\right]/\sqrt{2}$ (b) $D_s^*\bar{D}^{*}$.
		}\label{fig: Zcs Dis}
	\end{figure}

	In Refs. \cite{Maiani2021, Meng2021}, the $Z_{cs}(3985)$ and
	$Z_{cs}(4000)$ states are discussed as the SU(3)-flavor partners of
	$Z_c(3900)$, with their neutral nonstrange members having opposite C
	parity. The authors suggest that the $Z_{cs}(4000)/Z_{cs}(3985)$
	state can be described as a pure molecular state composed of
	$(|D_s\bar{D}^*\rangle +/-|D_s^*\bar{D}\rangle)/\sqrt 2$.
	Furthermore, they also predicted the existence of a $D_s^*\bar{D}^*$
	molecular state, which is potentially supported by the recent work
	of the BESIII collaboration \cite{BESIIICollaboration2023}. These
	studies provide interesting insights into the nature and composition
	of the $Z_{cs}$ states under the molecule picture.

	However, an issue that remains unresolved is the significant
	difference in the widths between the $Z_{cs}(3985)$ and the
	$Z_{cs}(4000)$. To address this difference, we propose an
	alternative explanation where these two states are considered as two
	resonances associated with the same system, namely
	$(D_{s}\bar{D}^{*}+D_{s}^{*}\bar{D})/\sqrt{2}$. According to our
	proposal, the $Z_{cs}(3985)$ corresponds to the resonance with a
	narrower width, while the $Z_{cs}(4000)$ corresponds to the
	resonance with a broader width, as illustrated in Fig. \ref{fig: Zcs
		Dis}. This interpretation differs from the prevailing viewpoints in
	literature.

			\begin{figure}[bp]
		\subfigure[]{ \label{fig: Zcs 11}
			\includegraphics[width=180pt]{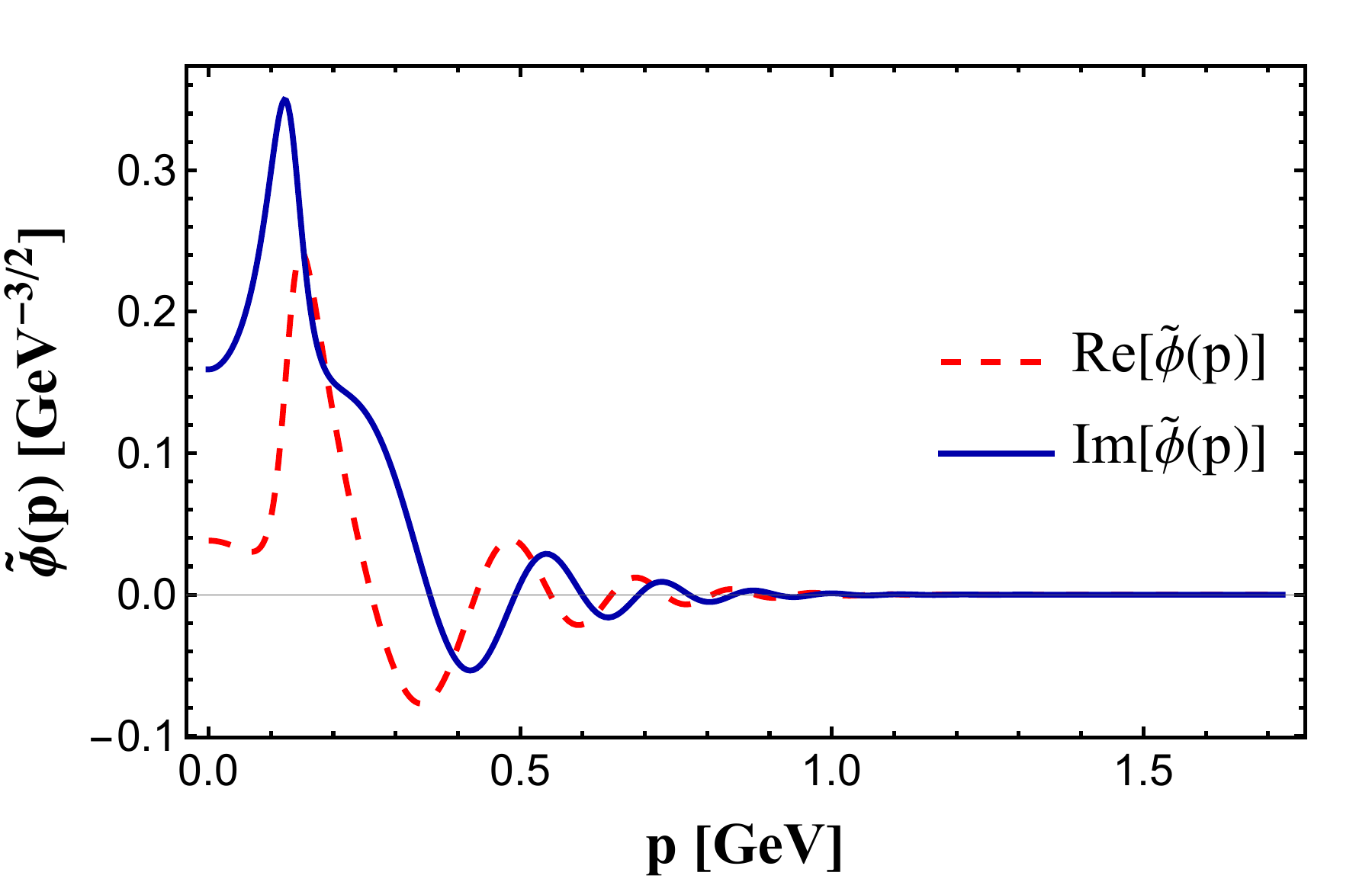}}\hspace{40pt}
		\subfigure[]{ \label{fig: Zcs 12}
			\includegraphics[width=180pt]{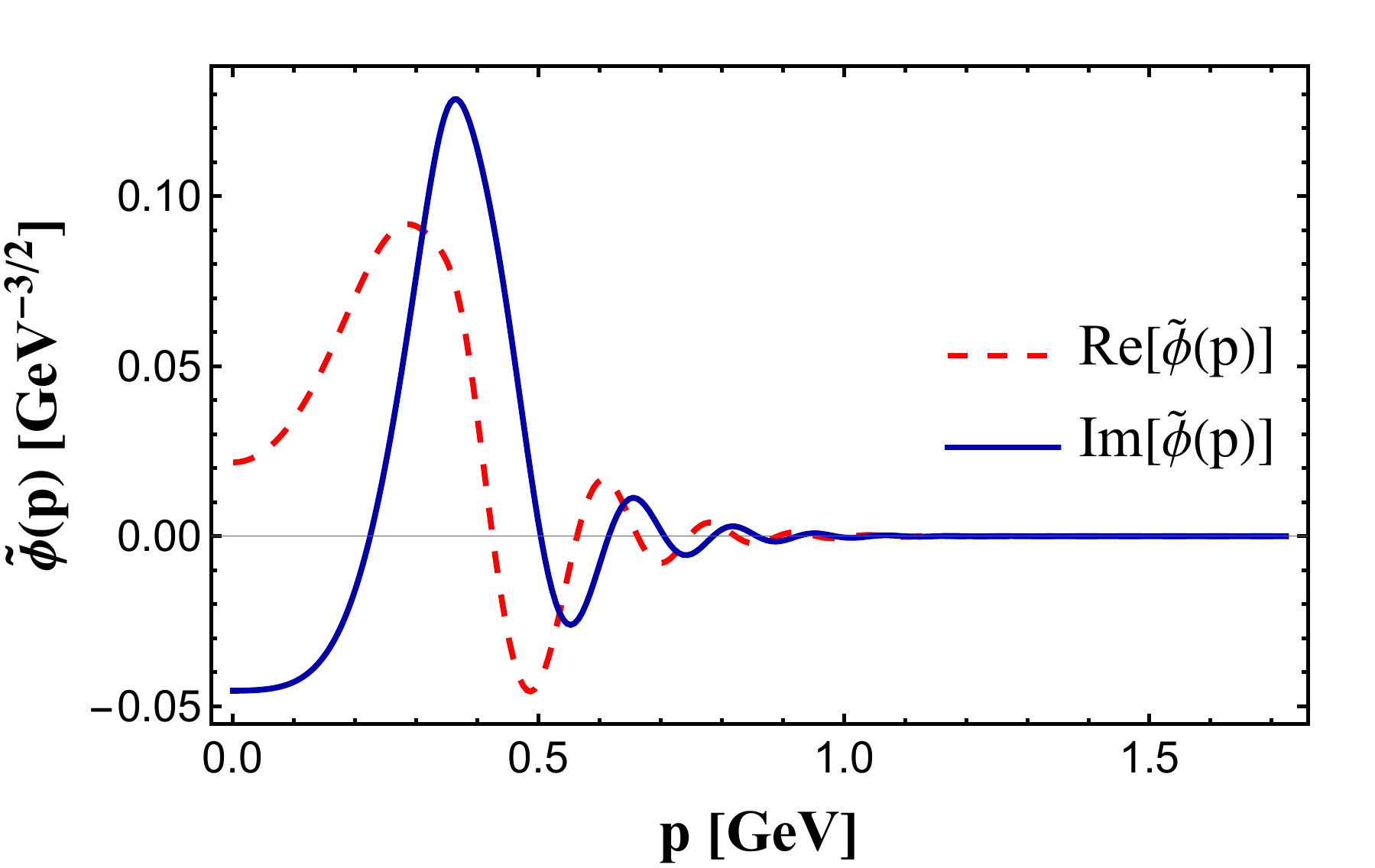}}
		\subfigure[]{ \label{fig: Zcs 21}
			\includegraphics[width=180pt]{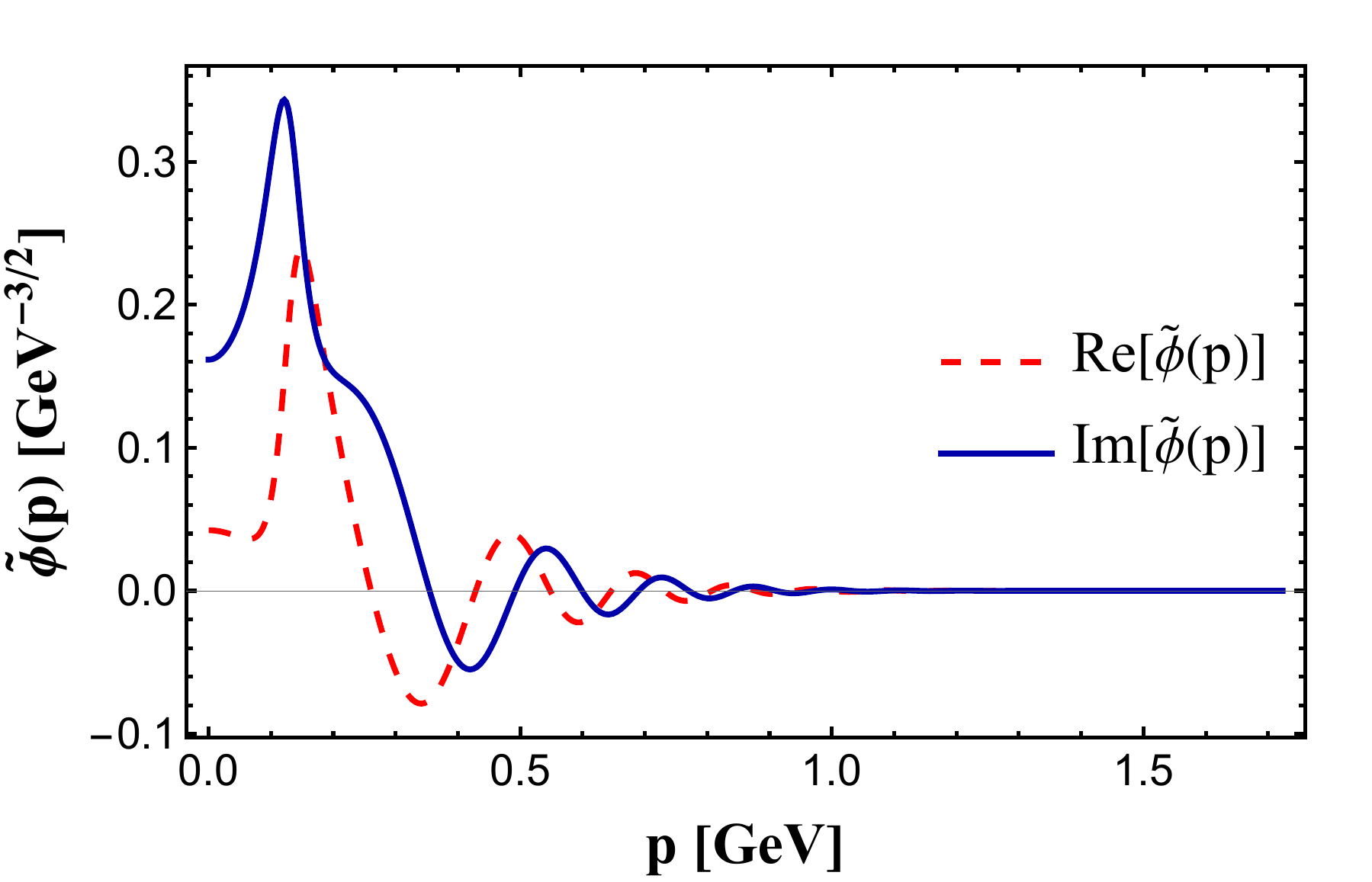}}\hspace{40pt}
		\subfigure[]{ \label{fig: Zcs 22}
			\includegraphics[width=180pt]{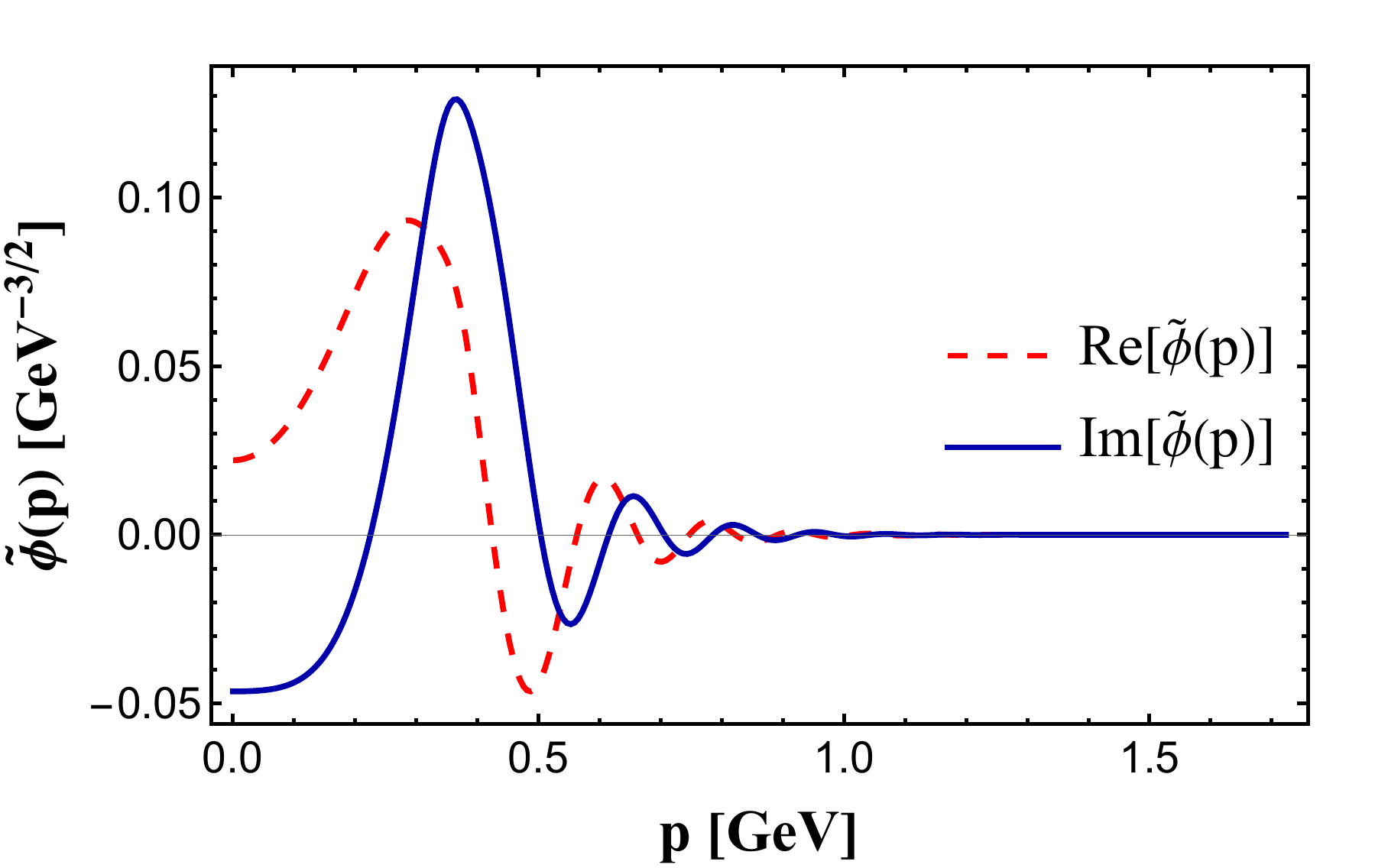}}
		\caption{\small The wave functions $\tilde{\phi}(p)$ of the
			$Z_cs$ state with the $I(J^P)=1/2(1^+)$. The rotation angle
			$\theta=40^\circ$ and the parameters $\Lambda=0.192$ GeV, $\tilde{C}_{s}=6.8\times10^2$ GeV$^{-2}$ and $C_{s}=-186.9\times10^2$ GeV$^{-4}$. The four diagrams correspond to: (a) pole 1 of $\left[D_s\bar{D}^{*}+D_s^*\bar{D}\right]/\sqrt{2}$ system (b) pole 2 of $\left[D_s\bar{D}^{*}+D_s^*\bar{D}\right]/\sqrt{2}$ system (c) pole 1 of $D_s^*\bar{D}^{*}$ system (d) pole 2 of $D_s^*\bar{D}^{*}$ system.}\label{fig: urobep}
	\end{figure}

	In the previous subsection \ref{sec:Zb and Zc}, we found that the
	$\left[D\bar{D}^{*}+D^*\bar{D}\right]/\sqrt{2}$ and $D^*\bar{D}^*$
	systems, associated with the two $Z_c$ states, exhibit similar
	outcomes due to the heavy quark spin symmetry. Thus, it is feasible
	to employ the same parameters for them. Following this scheme, we
	use the same parameters for both the
	$(D_{s}\bar{D}^{*}+D_{s}^{*}\bar{D})/\sqrt{2}$ and
	$D_{s}^*\bar{D}^{*}$ systems. By adopting the available experimental data of
	$Z_{cs}(3985)$ and $Z_{cs}(4000)$, we determine the central values
	and errors of $\Lambda$, $\tilde{C}{s}$ and $C_s$, and perform
	calculations for the $D_{s}^*\bar{D}^{*}$ system. The corresponding
	parameter values, masses, widths and sizes can be found in Table
	\ref{tab: results2}.

	In the framework of ChEFT, it is generally expected that the cutoff
	region should exceed the pion mass $m_\pi$ while not significantly
	exceeding $0.5$ GeV, as the higher-mass mesons ($\sigma$, $\rho$,
	$\omega$, etc.) are integrated out. Consequently, the $\Lambda$
	adopted in this study, $0.3\sim0.5$ GeV, is reasonable for the $Z_c$
	and $Z_b$ cases. However, in the case of $Z_{cs}$, the OZI
	suppression prohibits the contributions from either OPE or
	one-kaon-exchange. As a result, the contact term becomes the only
	interaction that needs to be considered. This can be viewed as
	effectively integrating out the pion and kaon fields. Therefore, we
	adopt a smaller value of $\Lambda\approx 0.2$ GeV for the $Z_{cs}$
	cases.

	According to the results in Table \ref{tab: results2}, the newly
	reported $Z_{cs}(4123)$ by BESIII collaboration
	\cite{BESIIICollaboration2023} could correspond to the narrower
	$D_{s}^*\bar{D}^{*}$ state, although the experimental width is yet
	to be confirmed. Its estimated mass is around 4125.2 MeV and width
	is approximately 13.2 MeV. Furthermore, the $Z_{cs}(4220)$ is
	anticipated to correspond to a broader resonance with its central
	values of the mass and width at 4152.7 MeV and 115.0 MeV. Indeed,
	the mass and width of $Z_{cs}(4220)$ both fall within the
	two-standard-deviation region of the experimental data.
	
	Furthermore, as shown in Fig. \ref{fig: urobep} and Table \ref{tab:
		results2}, the narrow (or broader) resonances exhibit remarkably
	similar wave functions and sizes.

	\section{Summary}\label{sec:summary}

	In this study, we employ the ChEFT to investigate the hidden-heavy
	tetraquark states with $I^G(J^{PC})=1^+(1^{+-})$ and the
	hidden-charm states with a strange quark with $I(J^{P})=1/2(1^{+})$
	in the molecule picture. 
	The couplings between the $S$-wave open-heavy channel and other channels, such as the $D$-wave channel, the $S$-wave channel with different constituents, and the hidden-heavy channels, are expected to be small.
	Therefore, we focus on the $S-$wave open-heavy
	single channels: $\left[D\bar{D}^{*}+D^*\bar{D}\right]/\sqrt{2}$,
	$D^*\bar{D}^*$, $\left[B\bar{B}^{*}+B^*\bar{B}\right]/\sqrt{2}$,
	$B^*\bar{B}^*$, $(D_{s}\bar{D}^{*}+D_{s}^{*}\bar{D})/\sqrt{2}$ and
	$D^*_s\bar{D}^*$.

	We employ the effective Lagrangians based on heavy quark symmetry
	and chiral symmetry, considering both contact and OPE diagrams. To
	investigate the possible resonances, we adopt the CSM to
	consistently analyze the bound states and resonances. In contrast to
	our previous work \cite{Cheng2022b,Cheng2023}, we perform the
	momentum space Schr\"odinger equation and discretize it using the
	Gaussian quadrature approach.

	In our investigation of the $Z_b$ system, we fit experimental data
	to extract resonance parameters within the molecule picture. With
	$\Lambda=0.510$ GeV, $\tilde{C}_s=0.48\times10^2$ GeV$^{-2}$ and
	$C_s=-5.4\times10^2$ GeV$^{-4}$, we obtain the mass and width values
	of 10606.9 MeV and 15.0 MeV for the
	$\left[B\bar{B}^{*}+B^*\bar{B}\right]/\sqrt{2}$ resonance, while
	10652.2 MeV and 14.8 MeV for the $B^*\bar{B}^*$ resonance. The RMS
	radii for these two resonances are both approximately $0.70-0.15i$
	fm. Similarly, we perform calculations for the $Z_c$ system in the
	$S-$wave $1^+(1^{+-})$ channels:
	$\left[D\bar{D}^{*}+D^*\bar{D}\right]/\sqrt{2}$ and $D^*\bar{D}^*$.
	Taking $\Lambda=0.300$ GeV, $\tilde{C}_s=2.86\times10^2$ GeV$^{-2}$
	and $C_s=-59.9\times10^2$ GeV$^{-4}$, we obtain the mass and width
	values of 3884.3 MeV and 26.0 MeV for the former resonance, while
	4025.8 MeV and 24.0 MeV for the latter resonance. The RMS radii for
	both resonances are around $1.20+0.13i$ fm. For the isovector
	$\left[D\bar{D}^{*}+D^*\bar{D}\right]/\sqrt{2}$ system, we also
	consider the influence of the three-body decay. However, the
	numerical results under the instantaneous approximation with $q_0=0$
	(as shown in the ``Inst'' row of Table \ref{tab: results}) are very
	close to the results of the ``Adopt'' row, indicating the minimal
	changes in the mass, width and size. Thus, we conclude that the
	2-body decay process dominates the width of this resonance.

	We consider the hidden-charm tetraquark states with a strange quark
	and propose that the $Z_{cs}(3985)$ and $Z_{cs}(4000)$ resonances
	correspond to the same channel
	$(D_{s}\bar{D}^{*}+D_{s}^{*}\bar{D})/\sqrt{2}$. Taking the data of
	$Z_{cs}(3985)$ and $Z_{cs}(4000)$ as input, we extract the central
	values and errors of the parameters $\Lambda$, $\tilde{C}_s$ and
	$C_s$. With $\Lambda=0.192$ GeV, $\tilde{C}_s=6.8\times10^2$
	GeV$^{-2}$ and $C_s=-186.9\times10^2$ GeV$^{-4}$, we obtain the
	mass and width values of 3982.4 MeV and 14.1 MeV for the
	$Z_{cs}(3985)$, while 4010.7 MeV and 119.6 MeV for the $Z_{cs}(4000)$.
	The corresponding RMS radii are $1.89+0.43i$ fm and $1.78+1.31i$ fm,
	respectively. For the $D^*_s\bar{D}^*$ system, we adopt the same
	parameters based on the heavy quark spin symmetry and also find two
	resonances. The narrower resonance has a mass of 4125.2 MeV and a
	width of 13.2 MeV, which nicely matches the observed $Z_{cs}(4123)$
	reported by the BESIII collaboration \cite{BESIIICollaboration2023}.
	Hence, we interpret it as the $Z_{cs}(4123)$, although the
	experimental width is yet to be confirmed. On the other hand, the
	broader resonance has a mass of 4152.7 MeV and a width of 115.0 MeV.
	We interpret it as the $Z_{cs}(4220)$ observed by the LHCb
	collaboration \cite{Aaij2021a}, as its mass and width fall within
	the two-standard-deviation region of the experimental data.

	In summary, we apply the ChEFT to investigate the $Z_b$, $Z_c$ and
	$Z_{cs}$ states. Our analysis suggests that the $Z_b(10610)$,
	$Z_b(10650)$, $Z_c(3900)$ and $Z_c(4020)$ can be interpreted as the
	molecular states formed by the $S-$wave $B\bar{B}^*$, $B^*\bar{B}^*$,
	$D\bar{D}^*$ and $D^*\bar{D}^*$ constituents, respectively. Although
	the $Z_{cs}(3985)$ and $Z_{cs}(4000)$ states exhibit a significant
	width difference, these two resonances may originate from the
	same $S-$wave channel $(D_{s}\bar{D}^{*}+D_{s}^{*}\bar{D})/\sqrt{2}$. We also
	find two resonances in the $D_s^*\bar{D}^*$ channel, which can be
	identified as the $Z_{cs}(4123)$ and $Z_{cs}(4220)$. Our results
	provide a prediction for the width of the $Z_{cs}(4123)$ that awaits
	experimental confirmation. Additionally, we offer a precise mass and
	width range for the $Z_{cs}(4220)$, which can guide future
	experimental searches for the hidden-charm tetraquarks with a
	strange quark.

	\acknowledgments{This research is supported by the National Science
		Foundation of China under Grants No. 11975033, No. 12070131001 and
		No. 12147168. The authors thank K. Chen, Y. Ma and B. Wang for
		helpful discussions.}
	
	\bibliographystyle{apsrev4-2}
	\bibliography{tot.bib}

\end{document}